 \documentclass[preprint,aps,prd,preprint]{revtex4-1}
\usepackage{amssymb}
\usepackage{amsmath}
\usepackage{graphicx}
\usepackage{dcolumn}
\usepackage{bm}
\usepackage{color}
\usepackage{datetime}
\usepackage{wasysym}
\usepackage[mathscr]{euscript} 
\usepackage{hyperref}
\usepackage{slashed}
\usepackage{cancel}
\usepackage{ulem}
\newcommand{\bfig}{\begin{figure}}
\newcommand{\efig}{\end{figure}}  
\usepackage[utf8]{inputenc} 
\newcommand{\bal}{\begin{align}}
\newcommand{\eal}{\end{align}}
\newcommand{\beqa}{\begin{eqnarray}}
\newcommand{\eeqa}{\end{eqnarray}}
\newcommand{\nnn}{\nonumber}
\newcommand{\nn}{\nonumber \\}
\newcommand{\es}{& = &}

\newcommand{\ps}{& + &}
\newcommand{\ms}{& - &}

\newcommand{\3}{\underline{\bf{3}}}
\newcommand{\s}{\underline{\bf{1}}}
\newcommand{\spr}{\underline{\bf{1}'}}
\newcommand{\sppr}{\underline{\bf{1}''}}

\newcommand{\ket}[1]{ {\left|{#1}\right\rangle} }
\newcommand{\bra}[1]{ {\left\langle{#1}\right|} }

\newcommand{\ba}{\begin{align}}
\newcommand{\ea}{\end{align}}

\begin{document}
% Use the \preprint command to place your local institutional report
% number in the upper righthand corner of the title page in preprint mode.
% Multiple \preprint commands are allowed.
% Use the 'preprintnumbers' class option to override journal defaults
% to display numbers if necessary
%\preprint{}

%Title of paper

% repeat the \author .. \affiliation  etc. as needed
% \email, \thanks, \homepage, \altaffiliation all apply to the current
% author. Explanatory text should go in the []'s, actual e-mail
% address or url should go in the {}'s for \email and \homepage.
% Please use the appropriate macro foreach each type of information

% \affiliation command applies to all authors since the last
% \affiliation command. The \affiliation command should follow the
% other information
% \affiliation can be followed by \email, \homepage, \thanks as well.
\DeclareGraphicsExtensions{.eps,.ps}

\title{Signatures of $A_4$ symmetry in the charged lepton flavour violating decays in a neutrino mass model}

\author{Rambabu Korrapati}
\email[Email Address: ]{rambabu@phy.iitb.ac.in}
\affiliation{Department of Physics, Indian Institute of Technology Bombay,
Mumbai 400076, India}
\author{Jai More}
\email[Email Address: ]{more.physics@gmail.com}
\affiliation{Department of Physics, Indian Institute of Technology Bombay,
Mumbai 400076, India}
\author{Ushak Rahaman}
\email[Email Address: ]{ushakr@uj.ac.za}
\affiliation{Centre for Astro-Particle Physics (CAPP) and Department of Physics, University of Johannesburg, PO Box 524, Auckland Park 2006, South Africa}
\author{S. Uma Sankar}
\email[Email Address: ]{uma@phy.iitb.ac.in}
\affiliation{Department of Physics, Indian Institute of Technology Bombay,
Mumbai 400076, India}
%%%%%%%%%%%%%%%%%%%%%%%%%%%%%%%%%%%%%%%%%%%%%%%%%%%%%
%Collaboration name if desired (requires use of superscriptaddress
%option in \documentclass). \noaffiliation is required (may also be
%used with the \author command).
%\collaboration can be followed by \email, \homepage, 
%\thanks as well.
%\collaboration{}
%\noaffiliation
%%%%%%%%%%%%%%%%%%%%%%%%%%%%%%%%%%%%%%%%%%%%%%%%%%%%%%%%%%%%%%%%%%%%%%%%
\date{\today}
%%%%%%%%%%%%%%%%%%%% abstract %%%%%%%%%%%%%%%%%%%%%%%%%%%%%%%%%%%%%%%%%%
\begin{abstract}
We study the charged lepton flavour violation in a popular neutrino mass model with $A_4$ 
discrete symmetry. This symmetry requires the presence of multiple Higgs doublets in the  
model and it also dictates the flavour violating Yukawa couplings of the additional neutral scalars 
of the model. Such couplings lead to the decays of the neutral mesons, the top quark and the  
$\tau$ lepton into charged leptons of different flavours at tree level. The $A_4$ symmetry of the 
model leads to certain characteristic signatures in these decays. We discuss these signatures and 
predict the rates for the most favourable charged lepton flavour violating modes. 
\end{abstract}
%%%%%%%%%%%%%%%%%%%%%%%%%%%%%%%%%%%%%%%%%%%%%%%%%%%%%%%%%%%%%%%%%%%%%%%%%
% insert suggested PACS numbers in braces on next line
%\pacs{14.60.Pq,14.60.Lm,13.15.+g}
%%%%%%%%%%%%%%%%%%%%%%%%%%%%%%%%%%%%%%%%%%%%%%%%%%%%%%%%%%%%%%%%%%%%%%%%%%
% insert suggested keywords - APS authors don't need to do this
%\keywords{Neutrino Mass Hierarchy, Long-Baseline Experiments}

\maketitle

\section{Introduction}
Neutrino oscillations provide the first hint of physics beyond the standard model. They also imply that the different lepton numbers, $L_e, L_\mu$ and $L_\tau$ are not conserved individually. Non-conservation of these quantum numbers opens up the possibility of flavour non-conservation in the charged lepton sector also. That is, decays such as $K_L\to \mu \,e$, $B_d, B_s \to \ell_1^+\, \ell_2^- $ and other flavour violating decays of heavy quarks and leptons should be possible. Various experiments have been searching for signals of charged lepton flavour violation during the past two decades.

Neutrino oscillations arise because neutrino flavour eigenstates are linear combinations of mass eigenstates. It is this mismatch which leads to flavour violation in the lepton sector.
To get the full picture of lepton flavour violation, we need a full-fledged theory of neutrino masses. Given such a theory, it is possible to establish connections between flavour violations in the neutrino sector and in the charged lepton sector. At present, there are many popular models of neutrino mass. Different models predict different values for the charged lepton violating decays, depending on the details of the model. For a review of charged lepton flavour violations in various popular neutrino mass model, see~\cite{Mohapatra:2005wg}.

The relation between neutrino flavour eigenstates and mass eigenstates is described by the unitary matrix called the PMNS (Pontecorvo-Maki-Nakagawa-Sakata) matrix. This matrix is parametrized in terms of three mixing angles, $\theta_{12},\, \theta_{13}$
and $\theta_{23}$ and a CP violating phase $\delta_{CP}$, in analogy to the CKM matrix of the quark sector. Neutrino oscillation data show that $\sin^2 \, \theta_{12}\approx \frac1{3}$, $\sin^2 \,\theta_{13}\ll 1$, and $\sin^2 \,\theta_{23}\approx  \frac1{2}$. The current long baseline experiments, T2K~\cite{Abe:2019vii} and NOvA~\cite{Acero:2019ksn}, are beginning to measure $\delta_{CP}$. However, the best fit values of the $\delta_{CP}$ preferred by the two experiments are widely different. T2K prefers $\delta_{CP}$ value close to maximal violation ($\delta_{CP} \approx -90^0$) whereas NOvA prefers a value close to no CP violation ($\delta_{CP}\approx 0$).
%The current long baseline experiments, T2K~\cite{Abe:2019vii} and NOvA~\cite{Himmel2020}, are beginning to measure 
%$\delta_{CP}$. However, the best fit values of the $\delta_{CP}$ preferred by the two experiments are widely different. T2K prefers 
%$\delta_{CP}$ value of $\delta_{CP} \approx -110^0$ whereas NOvA prefers a value of  $\delta_{CP}\approx 140$. 

Various discrete symmetries were proposed to explain the pattern of neutrino mixings. The simplest of these is the $\mu \leftrightarrow \tau$ exchange symmetry~\cite{Harrison:2002et, Xing:2015fdg} which predicts $\theta_{13}=0$ and $\theta_{23}= 45^0$ with $\theta_{12}$ is left as a model dependent parameter. A number of popular models are based on the group $A_4$~\cite{Babu:2002dz,Ma:2001dn, Altarelli:2005yx, He:2006dk}
 which predict the mixing matrix to be of tri-bi-maximal (TBM) form~\cite{Harrison:2002er}: that is, $\sin^2 \theta_{13}=0$, $\sin^2 \theta_{23}=1/2$, $\sin^2 \theta_{12}=1/3$. A particular $A_4$ based model, proposed in ref.~\cite{He:2006dk}, obtains the TBM form of the PMNS matrix purely from symmetry and symmetry breaking considerations without any fine tuning of parameters. In 
ref.~\cite{Dev:2015dha}, it was shown that the introduction of a small perturbation in the Majorana mass matrix of the heavy right-chiral neutrinos in this model, can lead to both a realistic value of $\sin^2 \theta_{13}\approx 0.02$ and maximal CP violation.

In this article, we study the charged lepton flavour violation in the model of ref.~\cite{He:2006dk}. This model contains four $SU(2)$ Higgs doublets. These consist of an $A_4$ singlet $\phi_0$ and an $A_4$ triplet $\phi_i\,\, (i=1,2,3)$. In addition, there is also an $A_4$ triplet of scalars $\chi$ which are singlets under $SU(2)$. These multiple Higgs representations are required to form the PMNS matrix in the TBM form purely from symmetry considerations. The presence of multiple Higgs doublets, in general, leads to flavour changing Yukawa couplings (FCYC) at tree level. Such couplings can lead to observable branching ratios for decays such as $K_L\to \mu \,e$, 
$B_d, B_s \to \ell_1^+\, \ell_2^- $ and other possible charged lepton violating processes. In particular, these decays carry the 
characteristic signatures of $A_4$ symmetry. Constraints arising from the charged lepton flavour violating processes, 
on the extended scalar sectors due to flavour symmetries, were studied in~\cite{Adulpravitchai:2009gi}.
The charged lepton flavour violation in $B$ meson decays was studied 
in~\cite{Sahoo:2015pzk, Duraisamy:2016gsd} in the context of lepto-quark models and in~\cite{Heinrich:2018nip} in the 
context of a neutrino mass model with an $A_4$ triplet of isospin singlet scalars.

The paper is organized as follows: We give a brief outline of the model of ref.~\cite{He:2006dk} in section \ref{sec2} and study its Yukawa Lagrangian in detail in section \ref{sec3}.
We first consider the fields in $A_4$ eigenbasis and then transform the fermion fields into mass eigenbasis and isolate the FCYC terms of interest. Later we consider the Higgs potential and obtain the transformation matrix that gives us the Higgs mass eigenstates. Finally, we write the FCYC terms in terms of the mass eigenstates of the fermions and the Higgs bosons. In section \ref{sec4}, we consider the neutral meson decays into charged leptons of different flavours. In section \ref{sec5}, we discuss the lepton flavour violating decays, along with their $A_4$ signatures, of the top quark and the $\tau$ lepton. We briefly discuss the loop induced processes, $\mu \to e + \gamma$ and 
muon $(g-2)$ in section \ref{sec6} and present our conclusions in the last section.
%%%%%%%%%%%%%%%%%%%%%%%%%%
\section{Brief description of the model}\label{sec2}
%%%%%%%%%%%%%%%%%%%%%%%%%%
The charged fermion content of the $A_4$ model of ref.~\cite{He:2006dk} is the same as that of the SM with the same gauge quantum numbers. The model also contains three right-chiral neutrinos which form a triplet representation of $A_4$, but have no gauge quantum numbers. The three left chiral $SU(2)$ doublets of quarks $(Q_{iL},i=1,2,3)$ and leptons  $(D_{iL},i=1,2,3)$ are also assumed to form triplet representations of $A_4$. The $SU(2)$ singlet right chiral charged fermions have non-trivial transformation properties under $A_4$. The gauge and the $A_4$ quantum numbers of all the fermions are shown below:
\begin{equation}
\begin{array}{c}
Q_{iL}=  \left(
\begin{array}{c}
u_{iL}\\
d_{iL}
\end{array}
\right) \sim \left( 3,2,\frac{1}{3} \right) \left( \3 \right) \\
\\d_{1R} \oplus d_{2R} \oplus d_{3R} \sim \left( 3,1,-\frac{2}{3}
\right)\left(\s \oplus \spr \oplus \sppr \right)
\\
\\u_{1R} \oplus u_{2R} \oplus u_{3R} \sim \left( 3,1,\frac{4}{3} \right)\left(\s \oplus \spr \oplus \sppr \right) 
\end{array}
\quad\
\begin{array}{c}
D_{iL}=  \left(
\begin{array}{c}
\nu_{iL}\\ 
\ell_{iL}
\end{array}
\right)
\sim \left( 1,2,-1 \right) \left( \3 \right) \\
\\
\ell_{1R} \oplus \ell_{2R} \oplus \ell_{3R} \sim \left( 1,1,-2 \right)\left(\s
\oplus \spr \oplus \sppr \right)\\
\\
\nu_R \sim \left( 1,1,0 \right)\left( \3 \right). \\
\end{array}
\end{equation}
The Higgs field content of the model is dictated by the requirement that the PMNS matrix should be in the TBM form. To this end, three distinct Higgs field representations are introduced: (a) an $A_4$ triplet of $SU(2)$ doublets $\phi_i, (i=1,2,3)$, (b) an $A_4$ singlet of $SU(2)$ doublet $\phi_0$ and (c) an $A_4$ triplet of $SU(2)$ singlets $\chi_i, (i=1,2,3)$. The gauge and $A_4$ quantum numbers of these fields are shown below:
\begin{equation}
\phi_i =  \left(
\begin{array}{c}
\phi_{i}^{+}\\[2ex]
\phi_{i}^0
\end{array}
\right) \sim \left( 1,2,1 \right) \left( \3 \right),\quad
\phi_0 =  \left(
\begin{array}{c}
\phi_{0}^{+}\\
\phi_{0}^0
\end{array}
\right) \sim
\left( 1,2,1 \right) \left( \s \right),\quad 
\chi_i^0 \sim \left(
1,1,0 \right) \left( \3 \right).
\end{equation}
It is possible to write the Higgs potential in such a way that the Higgs fields have the following vacuum expectation values 
(VEV)~\cite{He:2006dk}: 
\begin{equation}
\langle \phi_1 \rangle =\langle \phi_2 \rangle =\langle \phi_3 \rangle = 
\left(
\begin{array}{c}
0\\[2ex]
v
\end{array}
\right), \,\,\, 
\langle \phi_0 \rangle = 
\left(
\begin{array}{c}
0\\[2ex]
v_0
\end{array}
\right), \,\,\, 
\langle \chi_i^0 \rangle = (0, w_2, 0).
\end{equation}
That is, all the three members of the $A_4$ triplet $\phi_i$ have the same VEV and only the second member of the $A_4$ triplet $\chi$ has a non-zero VEV. This arrangement of VEVs is crucial to obtain TBM form of PMNS matrix purely from symmetry considerations. 
%%%%%%%%%%%%%%%%%%%%%%%%%%%%%%
%%%%%%%%%%%%%%%%%%%%%%%%%%%%%%
 \section{Yukawa Lagrangian and charged fermions in mass eigenbasis}\label{sec3}
%%%%%%%%%%%%%%%%%%%%%%%%%%%%%%
%%%%%%%%%%%%%%%%%%%%%%%%%%%%%%
The Dirac mass terms for the fermions arise through the Yukawa interactions between the $SU(2)$ doublet Higgs fields and the fermion fields. Majorana masses for the neutrinos occur partly through bare mass terms and partly through Yukawa couplings of right-chiral neutrinos to the Higgs field $\chi_i^0$. 
The gauge and $A_4$ invariant Yukawa Lagrangian of this model, along with the bare Majorana mass terms, is given by \cite{He:2006dk, Grimus:2011fk}
\begin{eqnarray}\label{YLag}
 \mathcal{L}_{Yuk}
 \es -\Big[h_{1d} \,(\overline{Q}_{1L}\,\phi_1+\overline{Q}_{2L}\,\phi_2+\overline{Q}_{3L}\,\phi_3)\,d_{1R} 
+h_{2d}\,(\overline{Q}_{1L}\,\phi_1+\omega^2\overline{Q}_{2L}\,\phi_2+\omega\,\overline{Q}_{3L}\,\phi_3)\,d_{2R} \nn
 &&+h_{3d} \,(\overline{Q}_{1L}\,\phi_1+\omega\,\overline{Q}_{2L}\,\phi_2+\omega^2\,\overline{Q}_{3L}\,\phi_3)\,d_{3R}
 +h_{1u} \,(\overline{Q}_{1L}\,\tilde{\phi}_1+\overline{Q}_{2L}\,\tilde{\phi}_2+\overline{Q}_{3L}\,\tilde{\phi}_3)\,u_{1R} \nn
 &&+h_{2u} \,(\overline{Q}_{1L}\,\tilde{\phi}_1+\omega^2\,\overline{Q}_{2L}\,\tilde{\phi}_2+\omega\,
 \overline{Q}_{3L}\,\tilde{\phi}_3)\,u_{2R} 
 +h_{3u} \,(\overline{Q}_{1L}\,\tilde{\phi}_1+\omega\,\overline{Q}_{2L}\,\tilde{\phi}_2+\omega^2\,\overline{Q}_{3L}\,\tilde{\phi}_3)\,u_{3R}+ h.c. \Big]\nn
 &&
 -\Big[ h_{1\ell} (\overline{D}_{1L}\phi_1+\overline{D}_{2L}\phi_2+\overline{D}_{3L}\phi_3)\ell_{1R} 
 +h_{2\ell}(\overline{D}_{1L}\phi_1+\omega^2\overline{D}_{2L}\phi_2+\omega\overline{D}_{3L}\phi_3)\ell_{2R} \nn
 &&+ h_{3\ell}(\overline{D}_{1L}\phi_1+\omega\overline{D}_{2L}\phi_2+\omega^2\overline{D}_{3L}\phi_3)\ell_{3R}
 +h_0(\overline{D}_{1L}\,\nu_{1R}+\overline{D}_{2L}\,\nu_{2R}+\overline{D}_{3L}\,\nu_{3R})\tilde{\phi_0} + h.c. \Big]\nn
&&+ \frac1{2} \Big[M(\nu_{1R}^T\, C^{-1}\, \nu_{1R}+\nu_{2R}^T\, C^{-1}\, \nu_{2R}+\nu_{3R}^T\, C^{-1}\, \nu_{3R})+h.c.\Big]\nn
&&+ \frac1{2} \Big[h_\chi\, ((\chi_1 (\nu_{2R}^T\, C^{-1}\, \nu_{3R}+\nu_{3R}^T\, C^{-1}\, \nu_{2R})+
\chi_2 (\nu_{3R}^T\, C^{-1}\, \nu_{1R}+\nu_{1R}^T\, C^{-1}\, \nu_{3R})\nn
&&+ \chi_3 (\nu_{1R}^T\, C^{-1}\, \nu_{2R}+\nu_{2R}^T\, C^{-1}\, \nu_{1R}))+h.c.\Big],
 \end{eqnarray}
where $\tilde{\phi}_i = i \sigma_2 \phi^*_i$ and $\tilde{\phi}_0 = i \sigma_2 \phi_0^*$.
When the Higgs fields acquire their VEVs, this Lagrangian leads to mass matrices for charged fermions and the neutrinos of the form
\begin{equation}
- \bar{f}_L\, M_f\, f_R - \bar{\nu}_L\, M_D\, \nu_R +\frac1{2} \nu_R^T\, C^{-1} \,M_R \,\nu_R +h.c.
\end{equation}
Given the Higgs VEVs, we have $M_D = h_0 v_0 \mathbb{I}$ and 
\begin{equation}\label{MfMR}
M_f= \sqrt{3}\, v\, U_\omega^\dagger \left(
\begin{array}{c c c}
h_{1f} &~~ 0 ~~& 0\\[2ex]
0 & h_{2f}& 0\\[2ex]
0 & 0 & h_{3f}
\end{array}
\right)\,\mathbb{I},
\,\,
M_R= \left(
\begin{array}{c c c}
M & 0 & h_\chi w_2\\[2ex]
0 & M& 0\\[2ex]
h_\chi w_2 & 0 & M
\end{array}
\right),
\end{equation}
where $f=(u, d, \ell)$. 
The matrix $U_\omega$ is given by
\begin{equation}
 U_\omega = \frac{1}{\sqrt{3}}\left(
\begin{array}{c c c}
1 & ~~1~~ & 1\\[2ex]
1 & \omega & \omega^2\\[2ex]
1 & \omega^2 & \omega
\end{array}
\right),
\end{equation}
where $\omega$ is the cube root of unity. 
From eq.~(\ref{MfMR}), we note that the matrix $M_f$ is transformed to mass eigenbasis by making the unitary transformation $U_\omega$ on the left-chiral charged fermions $f_{iL}$ but leaving the corresponding right-chiral fields untouched.
In the case of charged fermions, we have the following relations between the Yukawa couplings and mass eigenvalues 
\beqa\label{lepmass}
h_{1f}=\frac1{\sqrt{3}}\frac{m_{1f}}{v},\,\, h_{2f}=\frac1{\sqrt{3}}\frac{m_{2f}}{v},\,\,h_{3f}=\frac1{\sqrt{3}}\frac{m_{3f}}{v},
\eeqa
where
$m_{1f}, m_{2f}$ and $m_{3f}$ are the masses of the first, second and third generation particles respectively.
Given that $m_{3f} \gg m_{2f} \gg m_{1f}$, we have 
\beqa\label{YCRel}
h_{3f} \gg h_{2f} \gg h_{1f}.
\eeqa
For charged leptons, the relation between the $A_4$ eigenstates and the mass eigenstates is
\beqa\label{Llepstate}
\ell_{1L}=\frac1{\sqrt{3}}(e_L+\mu_L+\tau_L),\,
\ell_{2L}=\frac1{\sqrt{3}}(e_L+\omega^2\,\mu_L+\omega\,\tau_L),\,
\ell_{3L}=\frac1{\sqrt{3}}(e_L+\omega\,\mu_L+\omega^2\,\tau_L).\nn
\eeqa
Relations similar to eq.~(\ref{Llepstate}) can be written for both up-type and down-type quarks. Since the same matrix $U_\omega$ transforms both up-type and down-type quark fields into their mass eigenstates, the CKM matrix $V_{CKM}=U_\omega^\dagger U_\omega =\mathbb{I}$. It is expected that radiative corrections can generate appropriate non-diagonal elements of this 
matrix~\cite{He:2006dk}. 

The diagonalizing matrix of $M_R$ is 
\begin{equation}
 U_\nu = \frac{1}{\sqrt{2}}\left(
\begin{array}{c c c}
1 & ~~~~0~~~&   -1\\[2ex]
0 & \sqrt{2} & 0\\[2ex]
1 & 0 & 1
\end{array}
\right),
\end{equation}
which leads to the PMNS matrix
\begin{equation}
U = U_\omega \,U_\nu = {\rm diag}(1, \omega, \omega^2)\, U_{\rm TBM}\, {\rm diag}(1, 1, - i),
\end{equation}
where the TBM form is 
\begin{equation}
 U_{TBM} = \left(
\begin{array}{c c c}
\frac2{\sqrt{6}} & ~~~~\frac1{\sqrt{3}}~~~&  0\\[2ex]
-\frac1{\sqrt{6}} &\frac1{\sqrt{3}} & -\frac1{\sqrt{2}}\\[2ex]
-\frac1{\sqrt{6}} &\frac1{\sqrt{3}}& \frac1{\sqrt{2}}
\end{array}
\right). 
\end{equation}
Here again, radiative corrections or other explicit $A_4$ breaking terms can lead to a phenomenologically viable form of the PMNS matrix with non-zero $\theta_{13}$ and $\delta_{CP}$.
  For example, in ref \cite{Dev:2015dha},
 $A_4$ symmetry was softly broken by unequal Majorana masses for $\nu_{iR}$. A simple adjustment of these masses leads to the correct value of $\theta_{13}$ and maximal CP violation.

In terms of the fermion mass eigenstates, the Yukawa Lagrangian can be written as:
\beqa\label{YCiFME}
\mathcal{L}_{Yuk}\es\mathcal{L}_{Yuk}^\ell +\mathcal{L}_{Yuk}^u+ \mathcal{L}_{Yuk}^d+\mathcal{L}_{Yuk}^\nu \nn
\rm{where}\nn
\mathcal{L}^\ell_{Yuk} \es-\frac{h_{1\ell}}{\sqrt{3}}\Big[
\left(\bar{e}_L+\bar{\mu}_L+\bar{\tau}_L\right)\phi_{1}^{0}+\left(\bar{e}_L+\omega\bar{\mu}_L+\omega^2\bar{\tau}_L\right)\phi_{2}^{0}+\left(\bar{e}_L+\omega^2 \bar{\mu}_L+\omega \bar{\tau}_L\right)\phi_{3}^{0}\Big]e_R\nn
&&-h_{1\ell}\Big[\bar{\nu}_{1L}\phi_{1}^{+}+\bar{\nu}_{2L}\phi_{2}^{+}+
\bar{\nu}_{3L}\phi_{3}^{+}\Big]e_R\nonumber \\
&&-\frac{h_{2\ell}}{\sqrt{3}}\Big[
\left(\bar{e}_L+\bar{\mu}_L+\bar{\tau}_L\right)\phi_{1}^{0}
+\omega^2\left(\bar{e}_L+\omega\bar{\mu}_L+\omega^2\bar{\tau}_L\right)\phi_{2}^{0}
+\omega\left(\bar{e}_L+\omega^2 \bar{\mu}_L+\omega \bar{\tau}_L\right)\phi_{3}^{0}\Big]\mu_R\nn
&&-h_{2\ell}\Big[\bar{\nu}_{1L}\phi_{1}^{+}+\bar{\nu}_{2L}\phi_{2}^{+}+
\bar{\nu}_{3L}\phi_{3}^{+}\Big]\mu_R\nonumber \\
&&-\frac{h_{3\ell}}{\sqrt{3}}\Big[
\left(\bar{e}_L+\bar{\mu}_L+\bar{\tau}_L\right)\phi_{1}^{0}
+\omega\left(\bar{e}_L+\omega\bar{\mu}_L+\omega^2\bar{\tau}_L\right)\phi_{2}^{0}
+\omega^2\left(\bar{e}_L+\omega^2 \bar{\mu}_L+\omega \bar{\tau}_L\right)\phi_{3}^{0}\Big]\tau_R\nn
&&-h_{3\ell}[\bar{\nu}_{1L}\phi_{1}^{+}+\bar{\nu}_{2L}\phi_{2}^{+}+
\bar{\nu}_{3L}\phi_{3}^{+}]\tau_R\nonumber \\
&&+\frac{h_0}{\sqrt{3}}\Big[(\bar{e}_L+\bar{\mu}_L+\bar{\tau}_L)\nu_{1R}+(\bar{e}_L+\omega \bar{\mu}_L+\omega^2 \bar{\tau}_L)\nu_{2R}+(\bar{e}_L+\omega^2 \bar{\mu}_L+\omega \bar{\tau}_L)\nu_{3R}\Big]\phi_{0}^{-}\nonumber \\
&& h.c.
\eeqa
$\mathcal{L}_{Yuk}^u$ and $\mathcal{L}_{Yuk}^d$ will have a similar structure but without the terms involving $h_0$. We will not explicitly discuss $\mathcal{L}_{Yuk}^\nu$ because it has no role to play in our calculations. 

In this model, the light neutrino mass spectrum is predicted to be nearly degenerate. 
The present direct upper limit on light neutrino mass is $1.1$ eV~\cite{Schluter:2020gdr}. 
It is possible to satisfy this limit if the common Dirac mass $h_0\, v_0 \sim m_e$ (of the 
order of 1 MeV) and if the heavy Majorana mass $M_R \simeq 1$ TeV. For simplicity, we 
further assume that $v$ and $v_0$ are equal, implying $v_0 = v=v_{\rm SM}/2=86$ GeV. 
Therefore, the Yukawa couplings $h_0$ and $h_{1\ell}$ are much smaller than the other two 
Yukawa couplings, {\it i.e.} $h_0, \, h_{1\ell} \ll h_{2\ell} \ll \,h_{3\ell}$. 
%Relaxing the assumption $v_0 = v$ does not change any of the qualitative features 
%of the charged lepton flavour violation that is the main focus of this work.
 %
\subsection{Higgs Potential of the model and the Higgs mass eigenstates:}
The most general  $A_4$ symmetric Higgs potential, in terms of the different $A_4$ representations, 
can be written as the sum of several parts,
\beqa
\label{Higgspot}
V\es V(\phi_i)+V(\chi)+V(\phi_0)+V(\phi_i,\chi)+V(\phi_i,\phi_0)+V(\phi_0,\chi)+V(\phi_i,\chi,\phi_0).
\eeqa
In ref.~\cite{He:2006dk}, there is a detailed discussion on the minimization of the Higgs potential in this model. 
The first three terms in eq.~(\ref{Higgspot}) correspond to self interaction of the three Higgs multiplets while the remaining terms give the interactions between them. 
To identify the various Higgs mass eigenstates, we need to diagonalize the matrix  $\Big(\partial^2 V/ \partial s_i \partial s_j\Big)_{VEV}$, where $s_i,\, s_j$ are two generic Higgs fields in the model. The full calculation is algebraically cumbersome. Hence we make some simplifying assumptions. We are interested in flavour changing neutral interactions of charged leptons mediated by scalars, which arise only due to the Yukawa couplings of the $SU(2)$ Higgs doublets. The $SU(2)$ singlet Higgs $\chi$ has no role to play in such interactions. Therefore, for simplicity, we neglect the admixture of $SU(2)$ doublets and $SU(2)$ singlet in forming the mass eigenstates. Hence we drop the terms containing $\chi$ in the Higgs potential. We make a further simplification which makes the algebra easier to handle but retains all the features of charged lepton flavour violations that are the focus of our work. Among the quartic terms of the potential, we keep only the terms containing the combination $(\phi_1^2+\phi_2^2+\phi_3^2)$ and set all other coefficients to be zero. This approximation makes the Higgs potential CP conserving.

The simplified Higgs potential is:
\beqa
V(\phi_\alpha)\es \mu_1^2 (\phi_1^2+\phi_2^2+\phi_3^2)+\lambda_1\,(\phi_1^2+\phi_2^2+\phi_3^2)^2+ \mu_2^2\,\phi_0^2+\lambda_3 \,\phi_0^4+\lambda_4 (\phi_1^2+\phi_2^2+\phi_3^2)\,\phi_0^2,
\eeqa
where $\phi_{\alpha}^{2}=\phi^\dagger_\alpha \,\phi_\alpha\,\, (\alpha=0,1,2,3)$. The mass squared matrix is obtained from the potential by
\beqa
\mathcal{M}_{\alpha\,\beta}^2\es\frac{\partial^2\, V(\phi_\alpha)}{\partial\,\phi_\alpha^*\,\partial\,\phi_\beta}\Bigg|_{VEV}, \nnn
\eeqa 
whose explicit form is 
\beqa
\mathcal{M}^2=  
\left(
\begin{array}{cccc}
\mu_2^2+4\lambda_3 v_0^2+3\lambda_4\, v^2
& \lambda_4 \,v_0\,v\,
& \lambda_4 \,v_0\,v\,
& \lambda_4 \,v_0\,v\,\\
\lambda_4 \,v_0\,v
& \mu_1^2+ 8\lambda_1\,v^2+\lambda_4\,v_0^2
&2\lambda_1\,v^2
&2\lambda_1\,v^2\\
 \lambda_4 \,v_0\,v\,
&2\lambda_1\,v^2
&\mu_1^2+ 8\lambda_1\,v^2+\lambda_4\,v_0^2
&2\lambda_1\,v^2
\\
\lambda_4 \,v_0\,v\,
&2\lambda_1\,v^2
& 2\lambda_1\,v^2
&\mu_1^2+ 8 \lambda_1\,v^2+\lambda_4\,v_0^2
\end{array}
\right).\nn
\label{mass-mat}
\eeqa
From the assumptions we made, it follows that the $\mathcal{M}^2$ is a real symmetric matrix which is diagonalized by the following orthogonal matrix,
\beqa
U_H=
\left(\begin{array}{cccc}\label{UH}
\frac{x}{\sqrt{3+x^2}}  & \frac{y}{\sqrt{3+y^2}}  & 0& 0\\
\frac1{\sqrt{3+x^2}}  & \frac1{\sqrt{3+y^2}} & -\frac1{\sqrt{2}}&\frac1{\sqrt{6}}\\
\frac1{\sqrt{3+x^2}} & \frac1{\sqrt{3+y^2}} & \frac1{\sqrt{2}}& \frac1{\sqrt{6}}\\
\frac1{\sqrt{3+x^2}} & \frac1{\sqrt{3+y^2}}& 0& -\frac2{\sqrt{6}}
\end{array}
\right).
\eeqa
In eq.~(\ref{UH}), the parameters $x$ and $y$ are defined by 
\begin{eqnarray}
x,y&=&\frac{1}{b}(a \pm \sqrt{a^2+ 3 b^2}),
\label{defxy}
\end{eqnarray}
where 
\begin{eqnarray}\label{defab}
a \es (2\lambda_4-8\lambda_1)v^2+2\lambda_3v_{0}^{2},  \nonumber \\
b\es 2\lambda_4\,v\, v_{0}.
\end{eqnarray}
From eqs.~(\ref{defxy}) and~(\ref{defab}), we find that $xy=-3$, which guarantees the orthogonality of the 
first two columns of $U_H$ and of its rows. 
 We denote the mass eigenstates of the neutral scalars to be $\Phi_\alpha^0, \, (\alpha=0,1,2,3)$. It can be shown that 
the imaginary part of $\Phi_0^0$ becomes the Goldstone boson coupling to $Z^0$ and the real part of 
$\Phi_0^0$ has the same properties as the SM Higgs boson. This can be identified with the $125$ GeV Higgs 
boson observed by ATLAS~\cite{Aad:2012tfa} and CMS~\cite{Chatrchyan:2012ufa} experiments. This model 
contains three heavier complex neutral scalars which are denoted by $\Phi_i^0,\,( i=1,2,3)$. 
The diagonalization of the $\mathcal{M}^2$ matrix in eq.~(\ref{mass-mat}) leads to degenerate eigenvalues 
for the states $\Phi_2^0$ and $\Phi_3^0$. The relationship between mass eigenbasis and $A_4$ eigenbasis of 
the $SU(2)$ doublet scalars is given by:
\begin{equation}\label{massbasis}
\phi_\alpha^0= (U_H)_{\alpha\,\beta}\,\Phi_\beta^0.
\end{equation}

%%%%%%%%%%%%%%%%%%%%%%%%%%%%%%%
%%%%%%%%%%%%%%%%%%%%%%%%%%%%%%%
\subsection{Yukawa couplings in the mass eigenbasis of fermions and scalars}
%%%%%%%%%%%%%%%%%%%%%%%%%%%%%%
%%%%%%%%%%%%%%%%%%%%%%%%%%%%%%%
In this work, we are interested in tree level flavour changing couplings of charged fermions to neutral scalars. The terms in  
eq.~(\ref{YCiFME}), proportional to $h_0$, do not lead to such couplings. From now on, we concentrate on terms containing the couplings $h_{1f}$, $h_{2f}$ and $h_{3f}$. 
We take the relevant terms in eq.~(\ref{YCiFME}) and transform the scalars, which are in their $A_4$ eigenstates, into their mass eigenstates. With this transformation the Yukawa couplings are in the mass eigenbasis of both the fermions and the scalars.
\beqa
\label{YCwh2h3}
\mathcal{L}_{Yuk}^{\ell}\es
- \frac{h_{1\ell}}{\sqrt{3}}\Bigg[\left(\bar{e}_L+\bar{\mu}_L+\bar{\tau}_L\right)\left(\frac{1}{\sqrt{3+x^2}}\Phi_{0}^{0}+\frac{1}{\sqrt{3+y^2}}\Phi_{1}^{0}-\frac{1}{\sqrt{2}}\Phi_{2}^{0}+\frac{1}{\sqrt{6}}\Phi_{3}^{0}\right)\nonumber \\ 
\ps \left(\bar{e}_L+\omega\bar{\mu}_L+\omega^2\,\bar{\tau}_L\right)\left( \frac{1}{\sqrt{3+x^2}}\Phi_{0}^{0}+\frac{1}{\sqrt{3+y^2}}\Phi_{1}^{0}+\frac{1}{\sqrt{2}}\Phi_{2}^{0}+\frac{1}{\sqrt{6}}\Phi_{3}^{0} \right)\nonumber \\ 
\ps\left(\,\bar{e}_L+\omega^2\bar{\mu}_L+\omega\, \bar{\tau}_L\right)\left( \frac{1}{\sqrt{3+x^2}}\Phi_{0}^{0}+\frac{1}{\sqrt{3+y^2}}\Phi_{1}^{0}-\frac{2}{\sqrt{6}}\Phi_{3}^{0} \right)\Bigg]e_R\nonumber \\
\ms \frac{h_{2\ell}}{\sqrt{3}}\Bigg[\left(\bar{e}_L+\bar{\mu}_L+\bar{\tau}_L\right)\left(\frac{1}{\sqrt{3+x^2}}\Phi_{0}^{0}+\frac{1}{\sqrt{3+y^2}}\Phi_{1}^{0}-\frac{1}{\sqrt{2}}\Phi_{2}^{0}+\frac{1}{\sqrt{6}}\Phi_{3}^{0}\right)\nonumber \\ 
\ps \left(\omega^2 \bar{e}_L+\bar{\mu}_L+\omega\,\bar{\tau}_L\right)\left( \frac{1}{\sqrt{3+x^2}}\Phi_{0}^{0}+\frac{1}{\sqrt{3+y^2}}\Phi_{1}^{0}+\frac{1}{\sqrt{2}}\Phi_{2}^{0}+\frac{1}{\sqrt{6}}\Phi_{3}^{0} \right)\nonumber \\ 
\ps\left(\omega\,\bar{e}_L+\bar{\mu}_L+\omega^2\, \bar{\tau}_L\right)\left( \frac{1}{\sqrt{3+x^2}}\Phi_{0}^{0}+\frac{1}{\sqrt{3+y^2}}\Phi_{1}^{0}-\frac{2}{\sqrt{6}}\Phi_{3}^{0} \right)\Bigg]\mu_R\nonumber \\
\ms \frac{h_{3\ell}}{\sqrt{3}}\Bigg[\left(\bar{e}_L+\bar{\mu}_L+\bar{\tau}_L\right)\left(\frac{1}{\sqrt{3+x^2}}\Phi_{0}^{0}+\frac{1}{\sqrt{3+y^2}}\Phi_{1}^{0}-\frac{1}{\sqrt{2}}\Phi_{2}^{0}+\frac{1}{\sqrt{6}}\Phi_{3}^{0}\right)\nonumber \\ 
\ps \left(\omega \bar{e}_L+\omega^2\,\bar{\mu}_L+\,\bar{\tau}_L\right)\left( \frac{1}{\sqrt{3+x^2}}\Phi_{0}^{0}+\frac{1}{\sqrt{3+y^2}}\Phi_{1}^{0}+\frac{1}{\sqrt{2}}\Phi_{2}^{0}+\frac{1}{\sqrt{6}}\Phi_{3}^{0} \right)\nonumber \\ 
\ps\left(\omega^2\bar{e}_L+\omega\bar{\mu}_L+ \bar{\tau}_L\right)\left( \frac{1}{\sqrt{3+x^2}}\Phi_{0}^{0}+\frac{1}{\sqrt{3+y^2}}\Phi_{1}^{0}-\frac{2}{\sqrt{6}}\Phi_{3}^{0} \right)\Bigg]\tau_R+h.c.
\eeqa
The Yukawa couplings similar to eq.~(\ref{YCwh2h3}) can be written for down quark sector as well. The corresponding Yukawa couplings for the up quark sector are
 \beqa
 \label{YCwh2uh3u}
\mathcal{L}_{Yuk}^{u}\es
- \frac{h_{1u}}{\sqrt{3}}\Bigg[\left(\bar{u}_L+\bar{c}_L+\bar{t}_L\right)\left(\frac{1}{\sqrt{3+x^2}}\Phi_{0}^{0*}+\frac{1}{\sqrt{3+y^2}}\Phi_{1}^{0*}-\frac{1}{\sqrt{2}}\Phi_{2}^{0*}+\frac{1}{\sqrt{6}}\Phi_{3}^{0*}\right)\nonumber \\ 
\ps \left(\bar{u}_L+\omega\bar{c}_L+\omega^2\,\bar{t}_L\right)\left( \frac{1}{\sqrt{3+x^2}}\Phi_{0}^{0*}+\frac{1}{\sqrt{3+y^2}}\Phi_{1}^{0*}+\frac{1}{\sqrt{2}}\Phi_{2}^{0*}+\frac{1}{\sqrt{6}}\Phi_{3}^{0*} \right)\nonumber \\ 
\ps\left(\,\bar{u}_L+\omega^2\bar{c}_L+\omega\, \bar{t}_L\right)\left( \frac{1}{\sqrt{3+x^2}}\Phi_{0}^{0*}+\frac{1}{\sqrt{3+y^2}}\Phi_{1}^{0*}-\frac{2}{\sqrt{6}}\Phi_{3}^{0*} \right)\Bigg]u_R\nonumber \\
\ms \frac{h_{2u}}{\sqrt{3}}\Bigg[\left(\bar{u}_L+\bar{c}_L+\bar{t}_L\right)\left(\frac{1}{\sqrt{3+x^2}}\Phi_{0}^{0*}+\frac{1}{\sqrt{3+y^2}}\Phi_{1}^{0*}-\frac{1}{\sqrt{2}}\Phi_{2}^{0*}+\frac{1}{\sqrt{6}}\Phi_{3}^{0*}\right)\nonumber \\ 
\ps \left(\omega^2 \bar{u}_L+\bar{c}_L+\omega\,\bar{t}_L\right)\left( \frac{1}{\sqrt{3+x^2}}\Phi_{0}^{0*}+\frac{1}{\sqrt{3+y^2}}\Phi_{1}^{0*}+\frac{1}{\sqrt{2}}\Phi_{2}^{0*}+\frac{1}{\sqrt{6}}\Phi_{3}^{0*} \right)\nonumber \\ 
\ps\left(\omega\,\bar{u}_L+\bar{c}_L+\omega^2\, \bar{t}_L\right)\left( \frac{1}{\sqrt{3+x^2}}\Phi_{0}^{0*}+\frac{1}{\sqrt{3+y^2}}\Phi_{1}^{0*}-\frac{2}{\sqrt{6}}\Phi_{3}^{0*} \right)\Bigg]c_R\nonumber \\
\ms \frac{h_{3u}}{\sqrt{3}}\Bigg[\left(\bar{u}_L+\bar{c}_L+\bar{t}_L\right)\left(\frac{1}{\sqrt{3+x^2}}\Phi_{0}^{0*}+\frac{1}{\sqrt{3+y^2}}\Phi_{1}^{0*}-\frac{1}{\sqrt{2}}\Phi_{2}^{0*}+\frac{1}{\sqrt{6}}\Phi_{3}^{0*}\right)\nonumber \\ 
\ps \left(\omega \bar{u}_L+\omega^2\,\bar{c}_L+\,\bar{t}_L\right)\left( \frac{1}{\sqrt{3+x^2}}\Phi_{0}^{0*}+\frac{1}{\sqrt{3+y^2}}\Phi_{1}^{0*}+\frac{1}{\sqrt{2}}\Phi_{2}^{0*}+\frac{1}{\sqrt{6}}\Phi_{3}^{0*} \right)\nonumber \\ 
\ps\left(\omega^2\bar{u}_L+\omega\bar{c}_L+ \bar{t}_L\right)\left( \frac{1}{\sqrt{3+x^2}}\Phi_{0}^{0*}+\frac{1}{\sqrt{3+y^2}}\Phi_{1}^{0*}-\frac{2}{\sqrt{6}}\Phi_{3}^{0*} \right)\Bigg]t_R+h.c.
\eeqa
Note that, in eqs.~(\ref{YCwh2h3}) and~(\ref{YCwh2uh3u}), the couplings of $\Phi_2^0$ and $\Phi_3^0$ to charged fermions are purely flavour violating whereas those of $\Phi_0^0$ and $\Phi_1^0$ are purely flavour conserving. 
Hence, there are no tree level amplitudes for decays with flavour violation at only one vertex, such as $\mu \to e \, \bar{e}\,e$, $\tau \to \mu\, \bar{\mu} \,\mu$,  $\tau \to e \, \bar{e}\,e$, $K_L \to \mu^+\,\mu^-$, $B_d \to\mu^+\,\mu^-$ and $B_s \to\mu^+\,\mu^-$.
%$\tau \to \mu \,e \bar{e}$ $\tau \to e\, \mu\, \bar{\mu}$,

Before we consider the charged lepton flavour violating phenomenology of $\Phi_2^0$ and $\Phi_3^0$, let us consider
the limits on the masses the heavy neutral scalars. The couplings of $\Phi_0^0$ to fermions are expected to be flavour
diagonal because the real and imaginary parts of $\Phi_0^0$ are identified with the SM Higgs boson and the Goldstone boson 
coupling to $Z^0$ respectively. The flavour diagonal couplings of $\Phi_1^0$ make it a SM-like heavy Higgs boson
and it can be produced in proton-proton ($p-p)$ collisions by the same processes which produce the SM Higgs boson. 
Recently, the CMS experiment has set a lower limit on the mass of such scalar $m_{\Phi_1} \geq 1870$ 
GeV~\cite{Sirunyan:2019pqw}. Since, $\Phi_2^0$ and $\Phi_3^0$ have purely flavour violating couplings, they can not
be produced via gluon-gluon fusion in $p-p$ collisions. The dominant process for their production will be vector boson
fusion. At present, the lower limit on the masses of neutral heavy scalars produced via vector boson fusion is only 300 
GeV~\cite{ATLAS:2016nje}.
%%%%%%%%%%%%%%%%%%%%
\section{Lepton flavour violating decays of neutral mesons}\label{sec4}
%%%%%%%%%%%%%%%%%%%%
Decays of neutral mesons, made of down-type quarks, are studied in more detail compared
to neutral mesons made of up-type quarks. Here we limit ourselves to the decays of neutral
$K$, $B_d$ and $B_s$ mesons into charged leptons with flavour violation.
Consider the decay of the meson with quark content $\bar{q}_i q_j$ into the final state 
$\ell_m^+ \ell_n^-$, with $m \neq n$. Since $\Phi_2^0$ and $\Phi_3^0$ have flavour 
violating couplings to both quarks and to charged leptons, their exchange can mediate the above 
decays at tree level. The flavour changing couplings of these heavy neutral scalars can be written,
in generic form, as 
\begin{equation}
g^{ij} \bar{f}_{iL} f_{jR} \Phi_2^0 + \tilde{g}^{ij} \bar{f}_{iL} f_{jR} \Phi_3^0 
+ \left( g^{ji} \right)^* \bar{f}_{iR} f_{jL} \left( \Phi_2^0 \right)^* 
+ \left( \tilde{g}^{ji} \right)^* \bar{f}_{iR} f_{jL} \left( \Phi_3^0 \right)^*.
\end{equation}
From eq.~(\ref{YCwh2h3}), we find that 

\begin{eqnarray}
g^{ij} & = & \frac{h_j}{\sqrt{6}} (1 - \omega) \nonumber \\
\tilde{g}^{ij} & = &  \frac{h_j}{\sqrt{2}} \omega^2,
\label{oddvertices}
\end{eqnarray}
for the ``odd" permutations $(ij) = (21), (32), (13)$ and 
\begin{eqnarray}
g^{ij} & = & \frac{h_j}{\sqrt{6}} (1 - \omega^2) \nonumber \\
\tilde{g}^{ij} & = &  \frac{h_j}{\sqrt{2}} \omega,
\label{evenvertices}
\end{eqnarray}
for the ``even" permutations $(ij) = (12), (23), (31)$. 

\begin{figure}[t]
\centering
\includegraphics[scale=0.3]{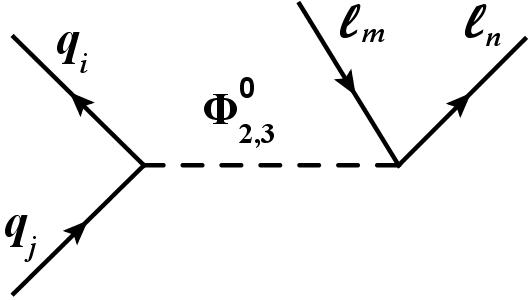}
\caption{\footnotesize{Tree level Feynman diagram for the $\bar{q}_i q_j \to \ell_m^+ \ell_n^-$
transition, mediated by $\Phi_2^0$ and $\Phi_3^0$.
}}
\label{figure1}
\end{figure}

The Feynman diagram for the transition $\bar{q}_i q_j \to \ell_m^+ \ell_n^-$ is
given fig.~\ref{figure1}. From the vertex factors given in eqs.~(\ref{oddvertices})
and~(\ref{evenvertices}), we find the four fermion amplitude to be 
\begin{eqnarray}
\left[ \frac{\left( g_q^{ij} \right) \left(g_\ell^{mn} \right)^*}{p^2 - m^2_{\Phi_2}} 
+ \frac{\left( \tilde{g}_q^{ij} \right) \left(\tilde{g}_\ell^{mn} \right)^*}{p^2 - m^2_{\Phi_3}} \right] 
<\ell_{mL}\bar{\ell}_{nR}|\bar{q}_{iL}q_{jR}> \nonumber \\
+\left[ \frac{\left( g_q^{ji} \right)^* \left(g_\ell^{nm} \right)}{p^2 - m^2_{\Phi_2}} 
+ \frac{\left( \tilde{g}_q^{ji} \right)^* \left(\tilde{g}_\ell^{nm} \right)}{p^2 - m^2_{\Phi_3}} \right] 
<\ell_{mR}\bar{\ell}_{nL}|\bar{q}_{iR}q_{jL}>,  
\label{decayamp}
\end{eqnarray}
where $g$ ($\tilde{g}$) correspond to the generic Yukawa coupling due to $\Phi_2^0 \, (\Phi_3^0)$ and
$p^2 \ll m^2_{\Phi_2}, m^2_{\Phi_3}$ is the momentum exchanged in the process. The coefficient
of the $\Phi_2^0$ exchange amplitude is {\bf equal} to that of the $\Phi_3^0$ exchange amplitude
if $(ij)$ and $(mn)$ are both even or both odd. If one is even and the other is odd, then the two 
coefficients are still of the same magnitude but of {\bf opposite} sign. For such cases, the two 
amplitudes exactly cancel each other in the limit $m_{\Phi_2} = m_{\Phi_3}$ that we consider here.

%Even if the masses of the two heavy neutral scalars are unequal, the decay rates of the cases with 
%even-odd pairing will be lower than decay rates of the cases with even-even or odd-odd pairings.
%For example, if $m_{\Phi_3} = 2 m_{\Phi_2}$, the decay rate of the favourable case will be about
%$3$ times larger than the decay rate of the unfavourable case. 

The net amplitude for the decay
$\bar{q}_i q_j \to \ell_m^+ \ell_n^-$ has two terms, one from the first line of eq.~(\ref{decayamp})
and one from the second line. The term from the first line has the coefficient $(h_j h_n)$ and the 
term from the second line has coefficient $(h_i h_m)$. Depending the values of $(ij)$ and $(mn)$,
one of these terms will dominate the other. 
The amplitudes due to $\Phi_2^0$ and $\Phi_3^0$ exchange add for the seven decays (and their 
charge conjugate decays) listed below.
\begin{itemize}
\item
$K^0 (\bar{s} d) \to \mu^+ e^-$ with coefficients $(h_{1d} h_{1\ell})$ and $(h_{2d} h_{2\ell})$
\item
$B_d^0 (\bar{b} d) \to e^+ \mu^-$ with coefficients $(h_{1d} h_{2\ell})$ and $(h_{3d} h_{1\ell})$
\item
$B_d^0 (\bar{b} d) \to \mu^+ \tau^-$ with coefficients $(h_{1d} h_{3\ell})$ and $(h_{3d} h_{2\ell})$
\item
$B_d^0 (\bar{b} d) \to \tau^+ e^-$ with coefficients $(h_{1d} h_{1\ell})$ and $(h_{3d} h_{3\ell})$.
\item
$B_s^0 (\bar{b} s) \to \mu^+ e^-$ with coefficients $(h_{2d} h_{1\ell})$ and $(h_{3d} h_{2\ell})$
\item
$B_s^0 (\bar{b} s) \to e^+ \tau^-$ with coefficients $(h_{2d} h_{3\ell})$ and $(h_{3d} h_{1\ell})$
\item
$B_s^0 (\bar{b} s) \to \tau^+ \mu^-$ with coefficients $(h_{2d} h_{2\ell})$ and $(h_{3d} h_{3\ell})$.
\end{itemize}
In the cases of the four decays, $K^0 \to \mu^+ e^-$, $B_d^0 \to \tau^+ e^-$, $B_s^0 \to \mu^+ e^-$ and
$B_s^0 \to \tau^+ \mu^-$ (and their charge conjugate modes), we have the product of the larger quark 
Yukawa coupling with the larger lepton Yukawa coupling. Hence these four decays are likely to have significant 
branching ratios in this model.

Before going into the details of the calculation, we would like to emphasize an important feature of
lepton flavour violation in this model. In the case of the decays $K^0 \to \mu^+ e^-$ and 
$\bar{K}^0 \to\mu^- e^+$, the amplitudes due to $\Phi_2^0$ and $\Phi_3^0$ exchange add but in
the case of the decays with charge conjugate final states,  $K^0 \to \mu^- e^+$ and 
$\bar{K}^0 \to \mu^+ e^-$, the two amplitudes cancel. Hence a neutral meson with the 
given flavour quantum numbers can decay only into a particular flavour combination of charged lepton pair 
but not into its charge conjugate pair. This charged lepton flavour selection is a {\bf signature} of the $A_4$ 
symmetry of the Yukawa couplings between the fermions and the scalar doublets in this model. But, such a 
signature will be difficult to observe experimentally in the case of neutral kaon decays because the physical 
decays observed are those of $K_L$ which contains roughly equal parts of $K^0$ and $\bar{K}^0$. Since the 
model predicts equal rates for $K^0 \to \mu^+ e^-$ and $\bar{K}^0 \to \mu^- e^+$, it predicts equal 
branching ratios for $K_L \to \mu^+ e^-$ and $K_L \to \mu^- e^+$. It may be possible to observe the above 
$A_4$ signature of charged lepton flavour selection in the leptonic decays $B_d$ mesons. Since $B_d^0 - \bar{B}^0_d$ 
are produced in pairs, it is possible to show that the final state $\tau^+ e^-$ occurred due to the decay of $B_d^0$,
rather than $\bar{B}_d^0$, by tagging the flavour of the $B$ meson on the other side.

Among the four favoured decays of neutral mesons to charged leptons discussed above, the experimental
upper bound on $\Gamma (K^0 \to \mu^+ e^-)$, which is easily related to the branching ratio of
$(K_L \to \mu^+ e^-)$, is the strongest. We use this mode to obtain a lower limit on $m_{\Phi}$, the common
mass of $\Phi_2^0$ and $\Phi_3^0$. Using this value of $m_\Phi$, we predict the branching ratios of the 
other three favoured decays in this model. From the expression in eq.~(\ref{decayamp}), we find the amplitude 
for $K^0 \to \mu^+ e^-$ to be
\begin{equation}
{\cal A} (K^0 \to \mu^+ e^-) = -\frac{h_{2d} h_{2\ell}}{4 m_\Phi^2} \bra{0} \bar{s} (1 - \gamma_5) d \ket{K^0}  
\bra{\mu^+ e^-} \bar{e} (1 + \gamma_5) \mu \ket{0}.
\label{K02mue}
\end{equation}
The decay rate for $K_L \to \mu^+ e^-$, due to this amplitude, is 
\begin{eqnarray}
\Gamma(K_L \to \mu^+ e^-) & = & \frac{1}{16 \pi} \left(1 - \frac{m_\mu^2}{m_K^2} \right)^2
\left( \frac{m_s}{m_s+m_d} \right)^2 \frac{m_\mu^2 f_K^2 m_K^5}{(3 v^2)^2 (2 m_\Phi)^4} \nonumber \\
& \approx & \frac{m_K}{16 \pi} \left( \frac{m_\mu}{\sqrt{3} v} \right)^2 \left( \frac{f_K}{\sqrt{3} v} \right)^2
\left( \frac{m_K}{2 m_\Phi} \right)^4 
\label{KL2mue}
\end{eqnarray}
where we set $(1 - m_\mu^2/m_K^2 \approx 1)$ and $m_s/(m_s+m_d) \approx 1$. In eq.~\ref{KL2mue},
$m_K (m_\mu)$ is the mass of the kaon (muon), $f_K$ is the kaon decay constant and $v = v_{\rm SM}/2$
is the common VEV of the four Higgs doublets (the $A_4$ singlet $\phi_0$ and the $A_4$ triplet
$\phi_i$). Substituting the appropriate values for the parameters in eq.~\ref{KL2mue} and comparing the 
resultant expression to the experimental upper bound $BR (K_L \to \mu^+ e^-) < 4.7 \times 
10^{-12}$~\cite{Ambrose:1998us}, we obtain the lower bound on $m_\Phi$ to be
\begin{equation}
m_\Phi \geq 750 \, {\rm GeV}.
\end{equation}

This bound satisfied the present experimental lower limit of 300 GeV~\cite{ATLAS:2016nje}. 
For the lowest allowed value of $m_\Phi$, the branching ratios of the other favoured modes 
are predicted to be 
\begin{eqnarray}
BR (B_d^0 \to \tau^+ e^-) & = & 8 \times 10^{-9},  \nonumber \\
BR (B_s^0 \to \mu^+ e^-) & = & 3.5 \times 10^{-11},  \nonumber \\
BR (B_s^0 \to \tau^+ \mu^-) & = & 8 \times 10^{-9}. 
\end{eqnarray}
The respective present experimental upper bounds on these branching ratios are 
$(3 \times 10^{-5})$~\cite{Aubert:2008cu}, $(5.4 \times 10^{-9})$~\cite{Aaij:2017cza} and
$(4.2 \times 10^{-5})$~\cite{Aaij:2019okb}.
%%%%%%%%%%%%%%%%
%%%%%%%%%%%%%%%%
\section{Signatures of $A_4$ symmetry in the decays of the $\tau$ lepton and the top quark}\label{sec5}
%%%%%%%%%%%%%
\subsection{Decays of $\tau$ lepton}
%%%%%%%%%%%%%%%

Important signatures of the $A_4$ symmetry occur in the decay of $\tau$ leptons into three charged leptons. 
From the Yukawa couplings in eq.~(\ref{YCwh2h3}), we can show that in the case of the decays with the same 
charge dileptons $\tau^- \to e^- e^- \mu^+$ and $\tau^- \to \mu^- \mu^- e^+$, the amplitudes to $\Phi_2^0$
and $\Phi_3^0$ exchange add but they cancel for the decays with opposite charge dileptons $\tau^- \to \mu^+ \mu^- e^-$ 
and $\tau^- \to \mu^- e^+ e^-$  (which, in principle, can occur with flavour violation at both vertices). This 
occurrence of same sign dileptons in $\tau^-$ decays is a very distinctive signature of the $A_4$ symmetry of the 
Yukawa couplings.

From the form of the vertices given in eqs.~(\ref{oddvertices}) and~(\ref{evenvertices}), we find that 
the amplitude for the decay $\tau^- \to \mu^- \mu^- e^+$ is proportional to $h_{2 \ell} h_{3 \ell}/
m_{\Phi}^2$. Since the Yukawa couplings to $\mu$ and $\tau$ are rather small, we find that 
the decay rate into this mode is quite small. We calculate the branching ratio of this mode to be 
$\simeq 10^{-14}$ for $m_{\Phi} = 750$ GeV.

The branching ratio for $\tau^- \to e^- e^- \mu^+$
will be smaller by four more orders of magnitude because the corresponding amplitude is proportional
to $h_{1 \ell} h_{3 \ell}/m_{\Phi}^2$. 
%%%%%%%%%%%%%%%%
\subsection{Decays of top quark}
%%%%%%%%%%%%%%%%%%
In this model, a number of flavour changing couplings of the top quark have amplitudes proportional to
the large top Yukawa coupling. Hence the branching ratios of the decays of the top quark, mediated by
$\Phi_2^0$ and $\Phi_3^0$ can be measurably large. The $A_4$ structure of the top quark couplings 
to $\Phi_2^0$ and $\Phi_3^0$ implies that the amplitudes for the decays $t \to (u,c) \ell_1^+ \ell_2^-$ 
have the following forms:
\begin{itemize}
\item
${\cal A}( t \to c \mu^+ e^-) \propto \frac{h_{3 u} h_{2 \ell}}{m_{\Phi}^2}$
\item
${\cal A} (t \to c \tau^+ \mu^-) \propto \frac{h_{3 u} h_{3 \ell}}{m_{\Phi}^2}$
\item
${\cal A}( t \to u \mu^+ \tau^-) \propto \frac{h_{3 u} h_{2 \ell}}{m_{\Phi}^2}$
\item
${\cal A} (t \to u \tau^+ e^-) \propto \frac{h_{3 u} h_{3 \ell}}{m_{\Phi}^2}$
\end{itemize}

In the above list, we omitted the two decays, whose amplitude is proportional to the electron
Yukawa coupling, which makes the branching ratio much smaller than those of the above modes.  
Also, note that the $A_4$ symmetry of the model is dictating the flavours and charges of the final
state leptons. Cancellation of the amplitudes due to $\Phi_2^0$ and $\Phi_3^0$ exchange means 
that the decays into final states with the lepton charges reversed can not occur. Since the dominant
production of the top quark is in the form of $t \bar{t}$ pairs, it is possible to identify the flavour
of the quark decaying into charged leptons of different flavours by tagging the flavour of the quark
on the opposite side. Thus, establishing the $A_4$ signature of the charged lepton flavour selection in
top decays is straight forward.

The two decays $t \to c \,\tau^+\, \mu^-$ and $t \to u\, \tau^+ \,e^-$ have the largest couplings 
possible and their branching ratios are $\simeq 10^{-9}$ for $m_{\Phi} = 750$ GeV. The 
branching ratios for the other two modes are $\simeq 5 \times 10^{-12}$. 

At present, the
upper bound on the branching ratio of the decays of top quark into charged leptons of different 
flavours is $2 \times 10^{-5}$~\cite{Gottardo:2018ptv}. Thus there is a possibility that the favourable two 
decays listed above can be observed in the next run of the Large Hadron Collider (LHC).

%%%%%%%%%%%%%%%%
\section{Other flavour violating processes}\label{sec6}
%%%%%%%%%%%%%

Neutral meson mixing usually provides the strongest possible constraints on tree level scalars
with flavour violating couplings. Such mixing involves quark flavour transitions $\bar{q}_i q_j \to \bar{q}_j q_i$.
Comparing it to $\bar{q}_i q_j \to \ell^+_m \ell^-_n$ transition, we note the permutation structure
involves the product $(ij)*(ji)$, which is {\bf always odd}. Hence the term due to $\Phi_3$ exchange 
exactly cancels the term due to $\Phi_2$ exchange and the neutral mixing is absent at tree level.

The exact cancellation of the neutral meson mixing due to the exchange of $\Phi_2^0$ and
$\Phi_3^0$ occurs only for the when the their masses are exactly equal. If these masses
are unequal, then the neutral meson mixing leads to a very strong constraint on the difference
of the two masses. The effective four fermion operator for $K^0 \to \bar{K}^0$ transition has the form
\begin{equation}
H_{\rm eff} (K^0 \to \bar{K}^0) = 2 \bar{s}_L d_R 
\left[ \frac{(g^{sd})(g^{ds})^*}{p^2 - m_{\Phi_2}^2} +
\frac{(\tilde{g}^{sd})(\tilde{g}^{ds})^*}{p^2 - m_{\Phi_3}^2} \right] \bar{s}_R d_L.
\end{equation}
If $m_{\Phi_2} = m_{\Phi_3} = m_\Phi$, the coefficient of this operator is
$$ 
\frac{(g^{sd})(g^{ds})^* + (\tilde{g}^{sd})(\tilde{g}^{ds})^*}{p^2 - m_\Phi^2} = 0. 
$$
For unequal masses, we write $m_{\Phi_2} = m_\Phi - \delta m$ and $m_{\Phi_3} = m_\Phi + \delta m$.
so that $\Phi_2^0$ and $\Phi_3^0$ are no longer degenerate. We assume that $\delta m \ll m_\Phi$ and 
keep terms which are first order in $\delta m/m_\Phi$. In this approximation, the four fermion operator becomes
\begin{equation}
H_{\rm eff} (K^0 \to \bar{K}^0) = -4 \frac{1}{m_\Phi^2} \frac{\delta m}{m_\Phi}
\left[ (g^{sd})(g^{ds})^* - (\tilde{g}^{sd})(\tilde{g}^{ds})^* \right] \bar{s}_L d_R  \bar{s}_R d_L.
\end{equation}
From the $g^{ij}$ and $\tilde{g}^{ij}$ couplings, we can show that 
$$
\left[ (g^{sd})(g^{ds})^* - (\tilde{g}^{sd})(\tilde{g}^{ds})^* \right] = - h_{1d} h_{2d} \omega.
$$ 
We find the neutral kaon mass difference to be 
\begin{equation}
\frac{\Delta M_K}{m_K} = \frac{1}{m_K^2} Re (\langle \bar{K^0} | H_{\rm eff} | K^0 \rangle ).
= \frac{1}{2} (h_{1d} h_{2d}) \frac{f_K^2}{m_\Phi^2} \frac{m_K^2}{(m_s + m_d)^2} 
\frac{\delta m}{m_\Phi}.
\end{equation}
For $m_\Phi = 750$ GeV, we get
$$
\frac{\Delta M_K}{m_K}(NP) =  10^{-14} \frac{\delta m}{m_\Phi}.
$$
Demanding that this should be less than the experimental value $7 \times 10^{-15}$ {\bf does not}
give a useful constraint on $(\delta m/m_\Phi)$.

We do, however, get a very strong constraint when we consider $\epsilon_K$, which characterizes
the CP violation in $K^0-\bar{K}^0$ mixing.  Since $Im (\omega) = \sqrt{3} Re (\omega)$, we get
\begin{eqnarray}
\epsilon_K (NP) & \approx & \frac{Im (M_{12}) (NP)}{\Delta m_K (exp)} = \frac{Im (M_{12}) (NP)}{m_K} 
\left( \frac{\Delta m_K (exp)}{m_K} \right)^{-1}  \nonumber \\
& = & \frac{\sqrt{3} \times 10^{-14} (\delta m/m_\Phi)}{7 \times 10^{-15}} \leq 
\epsilon_K (exp) = 2 \times 10^{-3}, \nonumber
\end{eqnarray}
which leads to the very strong constraint $(\delta m/m_\Phi) \leq 10^{-3}$.  Even if $\Phi_2^0$ and $\Phi_3^0$
have unequal masses, their splitting can not be more than a GeV or so. The amplitudes for the {\bf cancelling 
processes} are suppressed by $(\delta m/m_\Phi) \sim10^{-3}$ relative to the {\bf adding processes}. Hence,
the rates for the unfavoured processes, which are predicted to have zero rate in the case of exactly equal masses 
for $\Phi_2$ and $\Phi_3$, are smaller by a factor of $10^{-6}$ compared to the  rates of processes which were
discussed earlier in the paper.

\begin{figure}[t]
\includegraphics[scale=0.3]{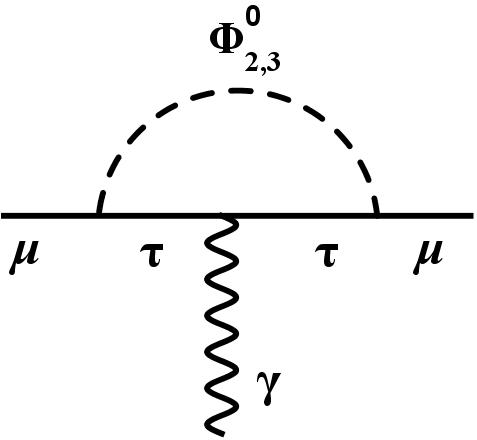}
\caption{\footnotesize{Feynman diagram for muon $(g-2)$ with $\Phi_2^0$ and $\Phi_3^0$ exchange.
}}
\label{figure2}
\end{figure}

Looking at the loop process $\mu \to e + \gamma$, we find that its amplitude also is a product
of even-odd permutations and hence vanishes in this model. On the other hand the muon $(g-2)$
operator has the product of odd-odd permutations for which the amplitudes due to $\Phi_2^0$ 
and $\Phi_3^0$ exchange add up. The corresponding Feynman diagram is shown in fig.~\ref{figure2} and the
amplitude is proportional to $h_{3 \ell}^2$, which is the largest leptonic Yukawa coupling. However,
the contribution of fig.~\ref{figure2} to muon $(g-2)$ is about $2 \times 10^{-14}$ for $m_\Phi
= 750$ GeV~\cite{Lavoura:2003xp}. This value is much smaller than the present discrepancy between the 
experiment and the theory. 

%%%%%%%%%%%%%%%
\section{Conclusions}\label{sec7}
%%%%%%%%%%%%%%%%
In this paper, we studied the charged lepton flavour violation in a neutrino mass model
with $A_4$ symmetry. This model has the attractive feature that it predicts the tri-bi-maximal
form of the neutrino mixing matrix purely from the symmetry considerations. The Yukawa 
couplings of the fermions to the multiple Higgs doublets of this model are guided by the $A_4$
symmetry. The flavour violating decays, mediated by heavy neutral scalars of this model,
carry signatures of the $A_4$ symmetry of the Yukawa couplings. 

We used a simplified form of the Higgs potential to obtain the Higgs mass eigenstates.
In this simplified form, two neutral scalars have only flavour conserving couplings and
two degenerate neutral scalars have only flavour violating couplings. In this situation,
the amplitudes for certain flavour violating transitions add and for certain other transitions
exactly cancel. Usually the neutral meson mixing and radiative charged lepton flavour violating 
decays provide the strongest constraints on the heavy neutral scalar masses. But there is {\bf no} 
contribution to these processes when the flavour violating neutral scalars of this model have 
exactly equal masses. If we assume a small splitting between the two masses, the CP violation
in kaon mixing constrains the splitting to be one part in a thousand. Thus, even if the 
neutral scalars have unequal masses, the rates of the unfavoured processes are about 
a million times smaller than the rates of the favoured processes.

We calculated the rates for a number favoured charged lepton flavour violating 
decays of neutral mesons, the top quark and the $\tau$ lepton in this work. Comparing the 
prediction of this model for the branching ratio of $K_L \to \mu^+ e^-$ to the present upper bound,
we derived a lower bound $m_{\Phi} \geq 750$ GeV, on the mass of the heavy neutral scalars
which have flavour changing couplings. We found a number of charged lepton flavour violating
decay modes of neutral $B$ mesons and the top quark which can be measured in the near future at 
LHC experiments or at Belle-II~\cite{Kou:2018nap}. 
\section*{Acknowledgments}
JM would like to thank the Department of Science and Technology (DST), Government of India, for financial support through Grant No. SR/WOS-A/PM-6/2019(G).

 %%%%%%%%%%%%%%%%%%%%%%%%%%%%%%%%%%%%%%%%%%%%%%%%%

\end{document}